\documentclass[letterpaper]{article} 
\usepackage{aaai20}  
\usepackage{times}  
\usepackage{helvet} 
\usepackage{courier}  
\usepackage[hyphens]{url}  
\usepackage{graphicx} 
\usepackage{graphicx}  
\usepackage{bm}
\usepackage{amsfonts}
\usepackage{amsmath}
\usepackage{multirow}
\usepackage{color}
\usepackage[table]{xcolor}
\usepackage{array}

\usepackage[T1]{fontenc}

\begin{document}

\title{AI and Core Electoral Processes: Mapping the Horizons}

\author{Deepak P$^1$ \hspace{0.2in} Stanley Simoes$^1$ \hspace{0.2in} Muiris MacCarthaigh$^2$ \\ \\
$^1$School of Electronics, Electrical Engg. and Computer Science, Queen's University Belfast, UK \\
$^2$School of History, Anthropology, Philosophy and Politics, Queen's University Belfast, UK \\
deepaksp@acm.org \hspace{0.2in} ssimoes01@qub.ac.uk \hspace{0.2in} M.MacCarthaigh@qub.ac.uk}

\maketitle

\begin{abstract}
Significant enthusiasm around AI uptake has been witnessed across societies globally. The electoral process -- the time, place and manner of elections within democratic nations -- has been among those very rare sectors in which AI has not penetrated much. Electoral management bodies in many countries have recently started exploring and deliberating over the use of AI in the electoral process. In this paper, we consider five representative avenues within the core electoral process which have potential for AI usage, and map the challenges involved in using AI within them. These five avenues are: voter list maintenance, determining polling booth locations, polling booth protection processes, voter authentication and video monitoring of elections. Within each of these avenues, we lay down the context, illustrate current or potential usage of AI, and discuss extant or potential ramifications of AI usage, and potential directions for mitigating risks while considering AI usage. We believe that the scant current usage of AI within electoral processes provides a very rare opportunity, that of being able to deliberate on the risks and mitigation possibilities, prior to real and widespread AI deployment. This paper is an attempt to map the horizons of risks and opportunities in using AI within the electoral processes and to help shape the debate around the topic. 
\end{abstract}


\section{Introduction}

Artificial Intelligence (AI) has rapidly spread across -- and in many cases, radically reshaped -- sectors as varied and diverse as medicine~\cite{rajpurkar2022ai} and transportation~\cite{iyer2021ai}. This has often been facilitated by a general propensity towards {\it tech exceptionalism} viz., the political viewpoint that tech-driven disruptions are largely positive~\cite{rosengrun2022ai}. That said, public sector adoption of AI has significantly trailed its uptake within the private sector, potentially due to considerations such as the effect of AI on human rights, political accountability, and its likely intensifying impact of existing power asymmetries~\cite{kuziemski2020ai}. There is a growing interest in considering the ethical, political, legal, policy and organizational challenges while using AI within critical sectors such as healthcare~\cite{sun2019mapping}. Some sectors in government such as policing has seen uptake of AI -- especially, predictive policing AI (e.g., PredPol\footnote{https://predpol.com/}) -- notwithstanding emerging understandings that paint a more nuanced, and often negative, picture of trade-offs between benefits and harms~\cite{mcdaniel2021predictive}. Yet, in the light of growing acceptance of AI as a significant technology, it has been forecasted that there would be enhanced adoption of AI within government, potentially freeing up one-third of public servants' time~\cite{berryhill2019hello}. 

In contrast to some other sectors of government, uptake of AI within the electoral process is very limited if not largely non-existent across most nations globally. We find it reasonable to assert that this situation is engendered by the negative impact that AI has been shown to have on public services, which have a high degree of implicit expectation to act fairly and responsibly. This viewpoint is also shared by The European Commission for Democracy through Law, who noted that digital technology can negatively affect the electoral process \cite[paragraph 7]{venice2020principles}, and in the abstract of a forthcoming article~\cite{bender}. Potential uptake of AI within recidivism prediction~\cite{propublica}, a task that sits on the fringes of judiciary, was largely responsible for attracting scholarly attention to biased AI operation and spawning research on fair AI, now a sub-discipline of bustling activity~\cite{chouldechova2020snapshot}. Deeper levels of AI uptake (e.g., robot judges) have been argued against~\cite{morison2019re}, and extant AI usage within China's court system has been viewed negatively~\cite{wang2020black}. The anti-poor consequences of AI usage within core governmental activity such as welfare application processing has been highlighted in popular literature~\cite{eubanks2018automating}. In short, these point to significant resistance towards AI usage within government. The uptake of AI within the media, the `fourth pillar' of democracy, has also met with significant criticism, especially in scenarios involving the overlap of AI-powered social media with electoral politics. These include the Facebook-Cambridge Analytica scandal~\cite{hinds2020wouldn}, and the impact of echo chambers on political fake news around elections~\cite{rhodes2022filter}.

We believe that these headwinds to AI within core government sectors, and thus the electoral process, creates a very rare space, one for deliberations on the pros and cons of AI usage within elections {\it prior} to their actual and substantive usage. We view the electoral process as being made up of {\it core} and {\it peripheral} functions, a conceptual distinction that we use to position this work. The core function is that of administration of the election, generally fulfilled by public {\it election bodies and authorities}\footnote{e.g., members of the Association of World Electoral Bodies http://aweb.org/eng/main.do}. On the other hand, the {\it peripheral} functions involve an ecosystem of private actors such as candidates who engage in campaigning, agencies that commission and conduct opinion polls, and media who report on the elections. We view the {\it core} and {\it periphery} as a structural and superficial dichotomy, and remark that peripheral functions (e.g., campaigning) are likely to be considered more important for elections, from a citizen's or civil society's point of view. The peripheral functions might be highly regulated by election bodies (e.g., financial and temporal limits on campaigning) to enable free and fair conduct of the core functions, and candidates' representatives may be allowed to have visibility of the core functions. The distinction between core and periphery, we note, aligns with the distinctions implied by {\it Article 1, Section 4} of the US constitution which has been interpreted~\cite{smith2013separation} as giving the state enhanced authority to regulate the {\it time, place and manner} (i.e., the core) of elections, and implying a separation of the campaign (i.e., periphery) and the state. Scholarly deliberations around AI and elections have largely focused on the peripheral functions, and have recently centered heavily on the impact of disinformation and personalized content on elections~\cite{stkepien2021ai}, an area that has been influenced significantly by AI (e.g., AI-powered fake news detection~\cite{al2022using,deepak2021data}). Some proposals in that space include deepening regulation~\cite{marsden2020platform} and understanding the multi-level challenges posed by {\it deepfakes}~\cite{whyte2020deepfake}. There has also been popular coverage of AI usage in other peripheral activities such as voter education and manifesto matching\footnote{https://www.forbes.com/sites/markminevich/2020/11/02/7-ways-ai-could-solve-all-of-our-election-woes-out-with-the-polls-in-with-the-ai-models/}. Our focus, in this paper, is on the {\it core} functions of the election process and to critically analyze the extant or potential role of AI within them. Our interest is in the electoral process used to choose representatives in representative democracies. 

Outside scholarly literature, the usage of AI in elections has seen some recent debates and discussions. In particular, the 2022 European Conference of Electoral Management Bodies\footnote{https://www.coe.int/en/web/electoral-management-bodies-conference/emb-2022}, organised by the Venice Commission (the Council of Europe's advisory body on constitutional matters including elections, referendums, and political parties\footnote{\url{https://www.venice.coe.int/WebForms/pages/?p=01_Presentation&lang=EN}}), had its theme as {\it Artificial Intelligence and Electoral Integrity}. The conclusions drawn from the conference\footnote{https://www.coe.int/en/web/electoral-management-bodies-conference/conclusions-2022} placed significant emphasis on the interference of tech giants on elections, noting concerns of personalization and selective exposure, microtargeting and voter turnout, synthetic AI-generated data vis-a-vis human oversight, and the issue of disinformation. We note that these significant concerns align more with the non-core electoral processes and are pertinent to the issue of regulating the information ecosystem. Further, a recent white paper \cite{heesen2021ai} and an associated spotlight article \cite{heesen2022ai} detail the risks associated with the use of AI in the information ecosystem during elections, and how it could reduce the agency of the voters and sway individual voting decisions.


In this paper, we consider five different avenues within the core electoral process within which AI currently plays or could potentially play a role in the near future. Where appropriate or necessary, we make the (potential) role of AI apparent by referencing related AI research. Our main goal is to map the ethical, social and political challenges that could be brought about by AI within those five avenues. While the overarching benefits of AI within those relate to efficiency improvements and savings in human labor, we observe that the cons of AI usage are of myriad types and intensities across the separate avenues. Accordingly, our critical analysis will place an enhanced focus on unpacking the negative ramifications of AI within the separate avenues. Given our focus on AI, non-AI automation within elections such as voting machines (e.g., India, Brazil) and online voting (e.g., Estonia) are not within our remit in this paper. We also do not concern ourselves with the different types of electoral systems (i.e. first-past-the-post, proportional representation, alternative vote, proportional representation - single transferable vote, etc), since that pertains only to the internals within the phase of vote counting. After the avenue-specific analyses focusing on the five chosen avenues, we will provide a high-level overview, and also briefly discuss other potential avenues. 

\begin{table*}
\resizebox{\textwidth}{!}{
\begin{tabular}{|c|c|c|c|}
    \hline
    \hline
    {\bf Avenue} & {\bf AI Usage} & {\bf Risks} & {\bf Pathways} \\
    \hline
    \hline
    \multirow{3}*{Voter List Maintenance} 
    & Heuristic-driven Approximations & Access-Integrity Trade-off Issues & Access-focused AI \\
    & Record Linkage & Biased AI & Reasonable Explanations \\
    & Outlier Detection & Overly Generalized AI & Local Scrutiny \\
    \hline
    \multirow{3}*{Polling Booth Locations} 
    & Drop Box Location Determination & Business Ethos & Plural Results \\
    & Facility Location & Volatility \& Finding Costs & Auditing AI \\
    & Clustering & Partisan Manipulation & Disadvantaged Voters \\
    \hline
    \multirow{3}*{Predicting Problem Booths} 
    & Predictive Policing & Systemic Racism & Transparency \\
    & Time Series Motifs & Aggravating Brutality & Statistical Rigor \\
    & & Feedback Loops & Fair AI\\
    \hline
    \multirow{4}*{Voter Authentication} 
    & Face Recognition & Race/Gender Bias & Alternatives \\
    & Biometrics & Unknown Biases & Bias Audits \\
    & & Voter Turnout & Designing for Edge Cases \\
    & & Surveillance and Misc. & \\
    \hline
    \multirow{3}*{Video Monitoring} 
    & Video-based Vote Counting & Electoral Integrity & Shallow Monitoring \\
    & Event Detection & Marginalized Communities & Open Data \\
    & Person Re-Identification & Undermining Other Monitoring & \\
    \hline
    \hline
\end{tabular}}
\caption{An overview of our avenue-specific analyses. \label{tab:avenuestable}}
\end{table*}


\section{AI usage Avenues within the \\ Core Electoral Process}

We consider the following five representative avenues of AI usage within the core electoral process. 

\begin{itemize}
    \item Voter List Maintenance and De-duplication
    \item Determining Polling Booth\footnote{Booths are used to refer to the location where a vote is cast, a predominant terminology in the Indian sub-continent. These are often referred to as polling stations in the UK, and polling places in the US.} Locations
    \item Vulnerability-based Polling Booth Protection
    \item Voter Authentication
    \item Video Monitoring of Electoral Fraud
\end{itemize}

While the first three avenues involve decision making {\it before} the actual elections, the following two relate to processes {\it during} the elections. While using the term AI, we use the contemporary interpretation that focuses on data-driven technologies based on machine learning and data analytics. Thus, traditional AI sub-disciplines such as planning and rule-based expert systems are outside our remit. 


\subsection{Avenue-specific Analyses}

Within each of the above avenues, our analysis will focus on critically and qualitatively evaluating the {\it context} as well as {\it extant or potential usage of AI technology}. This will be followed by our analysis of {\it potential risks}, and discussions on risk-mitigating or alternative {\it pathways}. An overview of our analyses appears in Table~\ref{tab:avenuestable}. 

\begin{figure}
  \includegraphics[width=\columnwidth]{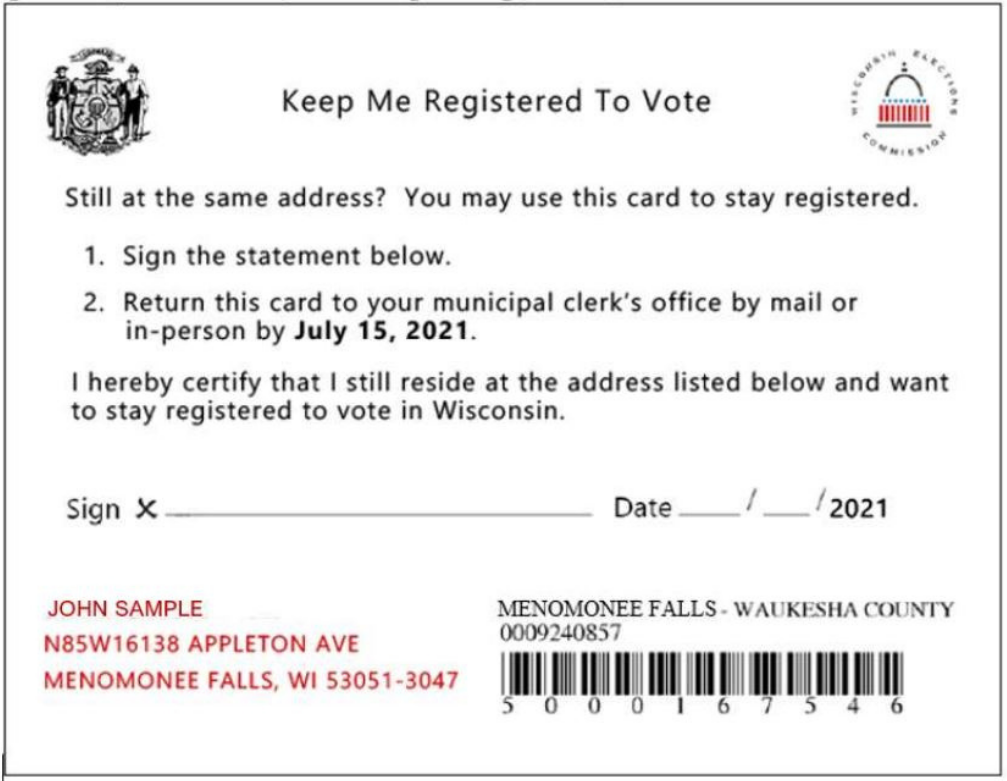}
  \caption{Sample of a card sent by Wisconsin to potentially inactive voters. The recipient has 30 days to take action and stay registered to vote. Pic: League of Women Voters (WI)}
  \label{fig:keep-me-registered}
\end{figure}

\section{Voter List Maintenance}

\subsection{The Context}

An up-to-date and error free voter list is critical to ensuring the integrity of elections~\cite{merivaki2020our}. Voter lists may be maintained at one of various levels of granularity viz., city/district, provincial or national levels. Further, not everybody may be eligible to vote in all elections. For example, in the Republic of Ireland, non-Irish citizens can vote in local elections but not in the presidential election, and EU citizens can vote in European Parliament elections but not national ones, or referenda\footnote{\url{https://www.gov.ie/en/service/a3e81-voting-in-ireland/}}. Thus, the voter lists may be specific to the elections in question. Whether there are multiple or single voters lists, there are overarching considerations to ensure that they are up to date. Voters need to be removed from voter lists if they have moved, changed citizenship, or in the event of their death. Analogously, incoming voters to a locality would need to be added to the voter list. In contrast to removal, the onus of making an application for inclusion in the list is usually on the incoming voter, and the application is accepted as long as set out conditions are met and the voter is not already registered. Periodically, there may also be a need to ensure that voters are not registered multiple times on the same voter list, or across different states' voter lists. 

\subsection{Extant or Potential AI Usage} 

The data entry corresponding to an ex-resident sticking around in a voter list could affect electoral integrity (especially, if a false vote is cast in their name), whereas a legitimate voter being left out of the list could reduce the confidence in the electoral process. Carefully balancing access and integrity is at the core of the task of voter list maintenance~\cite{merivaki2020our}. There has been little evidence of AI usage in voter list maintenance within scholarly or grey literature. There has, however, been a mention of the use of automated systems (e.g., name matching) to purge voter rolls in the US \cite{kayserbril2022algorithmic}, details of which would be available in a forthcoming article~\cite{bender}. In terms of non-traditional approaches to voter list maintenance, there have been cases of usage of administrative data (e.g., driving license data, postal service data) in Wisconsin~\cite{huber2021racial} to identify potential stale entries within the voter list, following which a postcard is sent to them to confirm that they are still resident in the address. Fig~\ref{fig:keep-me-registered} illustrates the format of a card sent out by Wisconsin to inactive voters, shifting the onus on the voter to then confirm that they are still resident at the address. We note that harvesting evidence from administrative sources and using patterns therein to determine potential staleness is a paradigm that can be significantly accelerated by AI. This is so since developing heuristics based on data patterns and harnessing them for faster decision making (often at some cost to accuracy) has been at the core of AI~\cite{batchelor2001algorithms}. In other words, this paradigm opens a channel for potential and substantive AI entry into the voter list maintenance process. 

We now outline streams of extant AI techniques which could be employed within voter list maintenance in the near future, based on our informed judgment. {\it First}, there has been a significant body of computing literature within the data management and analytics communities on the task of {\it record linkage}~\cite{christen2011survey}. These enable identifying duplicates or matching related records within a data source (e.g., voter list database) or across data sources (e.g., across voter list and driving license databases) using pattern-based heuristics. Identifying two similar entries in a voter list or identifying an entry in a voter list that matches with a driving license record for a different state could both be interpreted as pointing to potentially stale entries which need to be scrutinized further or removed. {\it Second}, techniques from the AI sub-discipline of {\it outlier detection}~\cite{wang2019progress} may be applied to identify `atypical' entries in voter lists, which may be flagged for manual scrutiny. Such usage is inherently problematic in that it is driven by the lack of appreciation of diversity among voters, but we suspect that the rampant usage of outlier detection in data-driven `smart' policing~\cite{yamin2020smart} (where similar considerations have been left unattended) would inevitably spill over to voter list maintenance. 

\subsection{Extant or Potential Ramifications} 

We consider potential or extant ramifications of AI usage in voter list maintenance at two levels viz., general and specific. 

At the general level, we draw the reader's attention again to the fundamental trade-off in voter list maintenance, that of balancing {\it access} and {\it integrity}~\cite{merivaki2020our}. While a focus on access would involve identifying the left-out voter, a focus on integrity would involve discovering fraudulent or stale entries in voter lists. The onus on the access issue is often left to individual voters to register themselves in time (and potentially, funded state campaigns to create public awareness), and thus, technological interventions on voter list maintenance are likely to focus on the integrity aspect. Data-driven AI is focused on leveraging available data sources such as voter lists in our scenario, and this makes it easier to conceptualize usage of AI to identify the fraudulent voter as opposed to the left-out voter who is absent in the voter list as it stands. In other words, deepening usage of AI in voter list maintenance is thus likely to lead to configurations that sit more towards the integrity side of the integrity-access trade-off. This may also be read within the backdrop of observations that extant AI usage has a dominant carceral-positive flavor~\cite{katz2020artificial}. 

At the specific level, we consider three representative issues. {\it First}, \cite{huber2021racial} observe that the paradigm of administrative data usage to identify fraudulent or stale voter list entries -- even through manual processes -- produce errors that engender a {\it `racial burden'} where errors are observed more for minority ethnicity. This implies that AI simply reproducing human decision making could itself be problematic, since it could produce the same kind of racial biases, but under an aura of technological legitimacy. {\it Second}, the usage of pattern-based heuristics to identify stale voter list entries would be inevitably faster and deeper with AI, as compared to manual processes. Such a data-led approach could lead to producing myriad kinds of biases as is often observed in data-driven AI systems~\cite{ntoutsi2020bias}. {\it Third}, there have been observations of significant local differences in voter list issues~\cite{merivaki2020our}. Such observations implicitly call for bespoke voter list maintenance techniques for specific localities to ensure effective working. We observe that this call stands in significant tension with the AI focus on generalizability, which has been observed as the second major driving value in machine learning research~\cite{birhane2022values}. In other words, normalizing AI for voter list maintenance could result in significant disadvantage for localities whose voter list maintenance issues are divergent from the predominant trends at the national or global level. 

\subsection{Pathways Forward} 

We discuss options that would sway AI adoption in voter list maintenance towards reasonable and low-risk directions. 

{\it First}, the access-integrity trade-off and the observations that current AI could be more applicable at the integrity end, points towards a research gap, that of {\it access-focused} AI. These would involve developing AI techniques to identify omissions in voter lists, perhaps using external sources such as administrative data (e.g., drivers licenses, postal service data). Such potential omissions could then be flagged to send targeted pamphlet-based requests to enroll in voter lists. These could help aid a more inclusive voter list to deepen the democratic process. {\it Second}, within the realm of AI usage towards identifying likely fraudulent or stale entries in voter lists, requiring usage of {\it explainable AI} that would generate {\it reasonable explanations} could be an important safeguard against biased operation. Obviously, what constitutes a reasonable explanation would necessarily need to be sourced from experts in voter list issues, and the technological challenge of developing AI that could adhere to such constraints would lead to new directions for AI research in the area. {\it Third}, each jurisdiction that maintains a voter list would need to analyze the appropriateness of off-the-shelf voter list maintenance AI for its own local context, given previously referenced observations on local differences in voter list issues~\cite{merivaki2020our}. This requires that AI techniques for voter list maintenance be transparent in their operation (this is complementary to producing explanations for each decision, as discussed above), so they can be scrutinized for applicability for each jurisdiction, and adapted easily to suit the specific conditions. 

\section{Polling Booth Location Determination}

\subsection{\bf The Context} 

\begin{figure}
  \includegraphics[width=\columnwidth]{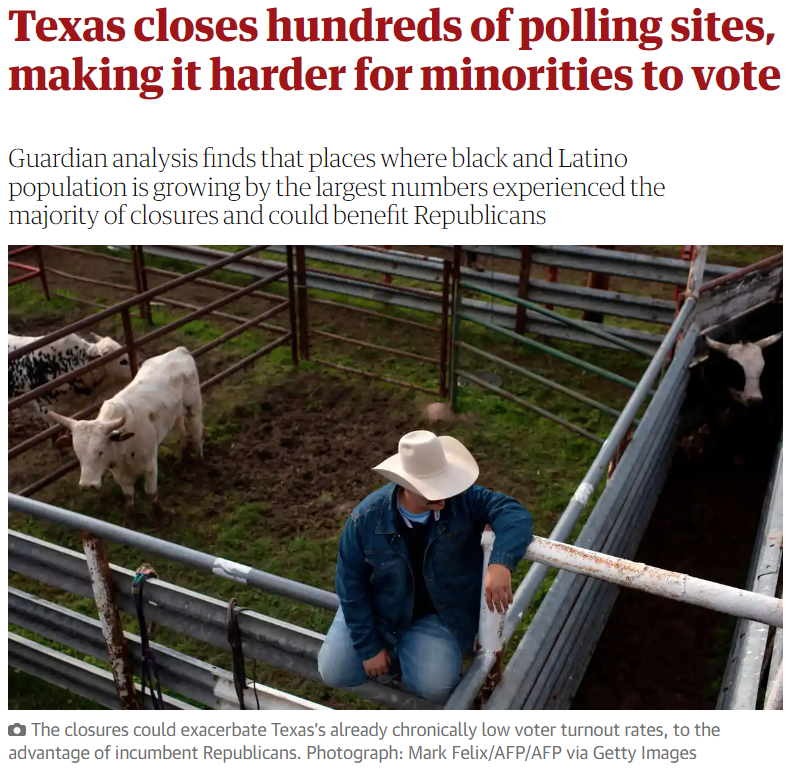}
  \caption{Screengrab from a March 2020 article in The Guardian on bias in polling booth closures in Texas.}
  \label{fig:texas-polling-booths}
\end{figure}

Polling booths, as locations where the voter exercises their right to choose, could be regarded as the most important location of an election. The importance of geographical locations have been long understood, not least due to concerns around gerrymandering, i.e., the practice of redrawing electoral district boundaries to the advantage of specific political parties~\cite{stephanopoulos2017causes}. Polling booth location determination is a finer-grained task, that of determining where individual voters within an electoral district should go to exercise their franchise. While most polling booth locations are fairly static over decades if not centuries, there has been evidence from political psychology that the nature of the polling booth (e.g., a church vs. a secular building) could influence the vote~\cite{rutchick2010deus}; this has been referred to as the polling place priming effect~\cite{blumenthal2011polling}. Further, the distance to the polling booth has been argued to have a nuanced relationship with voter turnout~\cite{garnett2021came}. All these suggest the importance of determining polling booths reasonably to ensure free and fair elections. Yet, it needs to be observed that identifying polling booth locations is often a simple decision-making task that needs to be undertaken very infrequently. For example, the legacy polling booths may be schools (or church halls, as common in the West) which continue usage as polling booths for generations. In the unusual event that a regularly used polling booth is discontinued, the proximal school (or church) may be a natural replacement. The main potential context of automating the polling booth determination process could be in locations that experience significant demographic churn due to inward or outward migration. Extensive inward migration might necessitate identifying newer booths for capacity considerations, and outward migration might require discontinuation of some legacy booths. Further, periodically, some jurisdictions may re-determine/change polling booths en masse viz., rationalization of polling booths in India\footnote{https://www.thehindu.com/news/cities/Hyderabad/rationalisation-of-polling-stations-ceo-seeks-time/article24609851.ece}, or consolidation in US which led to recent accusations of bias in Texas\footnote{https://www.theguardian.com/us-news/2020/mar/02/texas-polling-sites-closures-voting} (screengrab in Fig~\ref{fig:texas-polling-booths}) and Georgia\footnote{https://www.reuters.com/article/us-usa-election-georgia-idUSKCN1L51ZP}. 

\subsection{Extant or Potential AI Usage} 

Based on an extensive search, we found no evidence of AI usage in determining polling booth locations within extant literature. The only computing method we could locate in this space is a recent work on the {\it drop box location problem}~\cite{schmidt2022locating}, which considers computationally determining locations to place ballot drop-boxes, a recent addition to the voting infrastructure in the US, especially during the COVID era. The work proposes an integer program to address the task. The paucity of AI work on the polling booth location problem may be regarded as unsurprising since AI (or any kind of automation) is often prioritized in cases involving repetitive decision making (which is not quite the case for the polling booth location problem), to allow that the upfront investment may be justified over a reasonable period of time. 

\begin{figure}
  \includegraphics[width=\columnwidth]{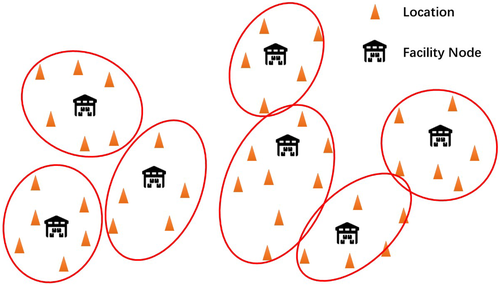}
  \caption{An illustration of clustering for facility location. For usage in polling booth determination, the locations could be locations of voters. Figure from~\cite{lin2021hierarchical}}
  \label{fig:clustering-facility-location}
\end{figure}

However, the technological readiness of AI for usage in polling booth determination may be regarded as moderate to high, owing to significant advances in related tasks. {\it First}, the classical task of {\it facility location}~\cite{celik2020comparative} may be observed to be highly allied with the task specification of polling booth determination. Facility location involves determining the locations for a (usually fixed) number of service centers (e.g., one of several types such as fire stations, post offices, cafes) in accordance with multiple (and potentially conflicting) criteria. For example, in the case of fire stations, the criteria to optimize for may include property value, population coverage and area coverage~\cite{farahani2010multiple}. A cafe chain may want to open a fixed number of new outlets to ensure accessibility to the maximum number of potential customers. When considering polling booths as facilities, area and population coverage are obviously pertinent criteria, in addition to other criteria such as public transport access. This specification of facility location naturally yields to multi-criteria optimization. {\it Second}, another classical task, that of {\it clustering}~\cite{xu2015comprehensive}, considers grouping objects into clusters/groups such that objects that are similar to one another be assigned to the same cluster with a high likelihood. The usage of clustering for facility location is most appropriate when the predominant assignment criterion is that of location-proximity, when similarity may be judged as inversely related to spatial distance. For the case of polling booths, voters may be clustered into geo-proximal clusters, beyond which each cluster (i.e., group of voters) may be allocated a polling booth; see Fig~\ref{fig:clustering-facility-location} for an illustration. Clustering algorithms have a long tradition that dates back to the 1960s~\cite{jain2010data}. The body of clustering literature is fairly versatile in that it includes algorithms that allow customized specifications for cluster shapes (e.g., spheres in k-means~\cite{macqueen2018some}), making them customizable in ways that facility location algorithms may not yield to. Vanilla clustering formulations seek to minimize the sum of distances between each object and its cluster representative; for the polling booth task, this would correspond to minimizing the cumulative distance that voters would need to travel to reach their polling booth. As may be obvious, minimizing the cumulative distance could still leave some voters with a significant distance to travel, to reach their polling booth. Recent advances in fair clustering~\cite{padmanabhan2020representativity,jung2020service,abbasi2021fair,stepanov2022fairness} consider ensuring that no objects (voters) are left too disadvantaged in terms of distance to their cluster representative (polling booth). This and other advances in fair clustering~\cite{mahabadi2020individual,kleindessner2020notion} have extended the clustering literature to align more with considerations that are applicable for polling booth location determination. 

\subsection{Extant or Potential Ramifications} 

We outline three potential risks based on our assessment of AI usage as exemplified above. {\it First}, facility location algorithms have traditionally considered scenarios within which at least one criteria is aligned with profitability or revenue. In fact,~\cite{celik2020comparative} explicitly suggest that the overarching trade-off in facility location is one between profitability and sustainability (of the business in question). The proposed technique for the drop box location problem~\cite{schmidt2022locating} also notably considers cost as a dominant criterion to optimize for. As an example, profitability can be improved by locating a facility (e.g., a cafe) in a location that is likely to attract more footfall (e.g., preferring a high-street location) relative to competition. However, polling booths do not operate in a competitive market since each voter is uniquely assigned to a polling booth, and the electoral body has a monopoly over the polling process. Thus, significant care must be used to ensure that business ethos does not creep into polling booth determination, and that off-the-shelf technology is customized enough to suit the task. {\it Second}, the availability of polling booth location determination technology may create an implicit urge to use it reasonably frequently, to extract value from the upfront cost of technology development or procurement. This may cause unfamiliar volatility in polling booth determination, with people seeing their assigned polling places change regularly. Such volatility in polling booth locations could increase the {\it finding costs}~\cite{brady2011turning} of polling places, and negatively affect voter turnout. {\it Third}, probably the most significant concern is that of technology enabling fine-grained gerrymandering, where polling locations (not electoral boundaries as in traditional gerrymandering) are manipulated for partisan objectives (we note here that there has been no evidence suggesting partisan manipulations of polling location yet~\cite{shepherd2021politics}). Yet, we draw attention to recent research~\cite{Fitzsimmons2020Selecting} proposing techniques for geographic manipulation of polling locations for partisan objectives. As an illustrative example, one of the tasks the paper considers is whether a set of polling locations can be chosen to ensure the victory of a specified candidate, under certain (arguably strong) assumptions. 




\subsection{Pathways Forward} 

We now consider risk-mitigating and progressive ways of using technology for polling booth determination to align with the best interests of free and fair elections. 

{\it First}, identifying a set of polling booth locations is a collective optimization task where several criteria are simultaneously optimized for. Given the result of the optimization, i.e., a set of suggested polling booth locations, a human expert (e.g., an election official) who considers one of the locations inappropriate would not be justified in altering it slightly, as we will explain. The judgment of inappropriateness of a particular polling booth location may stem from any of a variety of extrinsic reasons (e.g., proximity to a garbage dump, unavailability of adequate toilet facility etc.) that cannot be mathematically abstracted and fed into the optimization. The process of altering a specific polling booth location slightly is inappropriate since even a slight change in one polling booth may necessitate changes in other booth locations to ensure that the result stays truthful to the collective optimization at play. For the reader familiar with the task of centroid clustering (e.g., k-means~\cite{macqueen2018some}), a simpler collective optimization problem, the above scenario may be seen as analogous to the case where altering one cluster center could cause knock-on changes in other cluster centers and cluster allocations of even far flung objects. This inability to legitimately modify the AI's result even slightly could be seen as putting a human expert in a metaphorical straitjacket. Towards addressing this, AI techniques that produce a plethora of viable options may be developed, to allow the human to exercise due diligence and choose one from the viable options. {\it Second}, the advent of AI tools to enable technological manipulations of polling places~\cite{Fitzsimmons2020Selecting} calls for techniques to safeguard against such manipulations. Such safeguards can themselves be operationalized as auditing oriented AI which will enable discovery of partisan manipulations. There are, however, risks that such auditing AI could be used within adversarial learning frameworks~\cite{gui2021review} to enable the development of more refined AI-based malicious manipulations. {\it Third}, any form of polling booth location determination, whether AI-based or human-driven, would lead to a subset of disadvantaged voters. These could include voters who experience an awkward combination of physical disability and moderate to high distance to their assigned polling location. A novel stream of AI that identifies voters who are disadvantaged could aid efforts (e.g., prioritizing transportation help, or providing remote voting options) to foster more inclusive voting and offset the deficits of the chosen polling booth configurations. 






\section{Polling Booth Protection}

\subsection{The Context} 

Among the phenomena that threaten voters' exercise of voting franchise is {\it voter intimidation}, increasingly recognized as a global phenomenon (e.g., Guatemala~\cite{gonzalez2020carrots}, Russia~\cite{frye2019hitting}). Some jurisdictions have strong legal protections~\cite{woodruff2011wild}, but voter intimidation nevertheless remains existent. Other forms of infringement of voting rights include booth capturing~\cite{sharma1999booth} through violent means. These concerns have been used to justify preventive detention of potential criminals~\cite{verma2005policing} and designating specific polling places as {\it problem booths} to provide additional police protection\footnote{https://www.thehindu.com/news/cities/Kochi/276-problem-booths-in-district/article33273456.ece}. Such categorization and density gradients in polling booth police presence and patrolling could influence electoral calculus. In this section, we consider prioritization of polling booth protection as an avenue of potential AI usage. 

\subsection{Extant or Potential AI Usage} 

From our literature surveys, we find that concerns of intimidation and violence at polling places are highlighted largely within the Global South~\cite{electionviolence}, where AI penetration in the public sector has been traditionally low. This provides a backdrop into the non-existence (as far as we could assess) of bespoke AI techniques for polling booth protection prioritization. There has been emerging theoretical AI work on prioritizing protection of {\it `voter groups'} to ensure result stability in elections~\cite{Dey2019A,Dey2021A}, under conditions that we regard as quite synthetic; these may need much adaptation to be used within the task of prioritizing polling booths to protect. Our choice of identifying problematic polling booths as an avenue for AI usage is strongly motivated by observations of burgeoning AI development in a task that is very similar in spirit, that of {\it hot spot policing} that seeks to identify crime-prone locations, which we describe below.

\begin{figure}
  \includegraphics[width=\columnwidth]{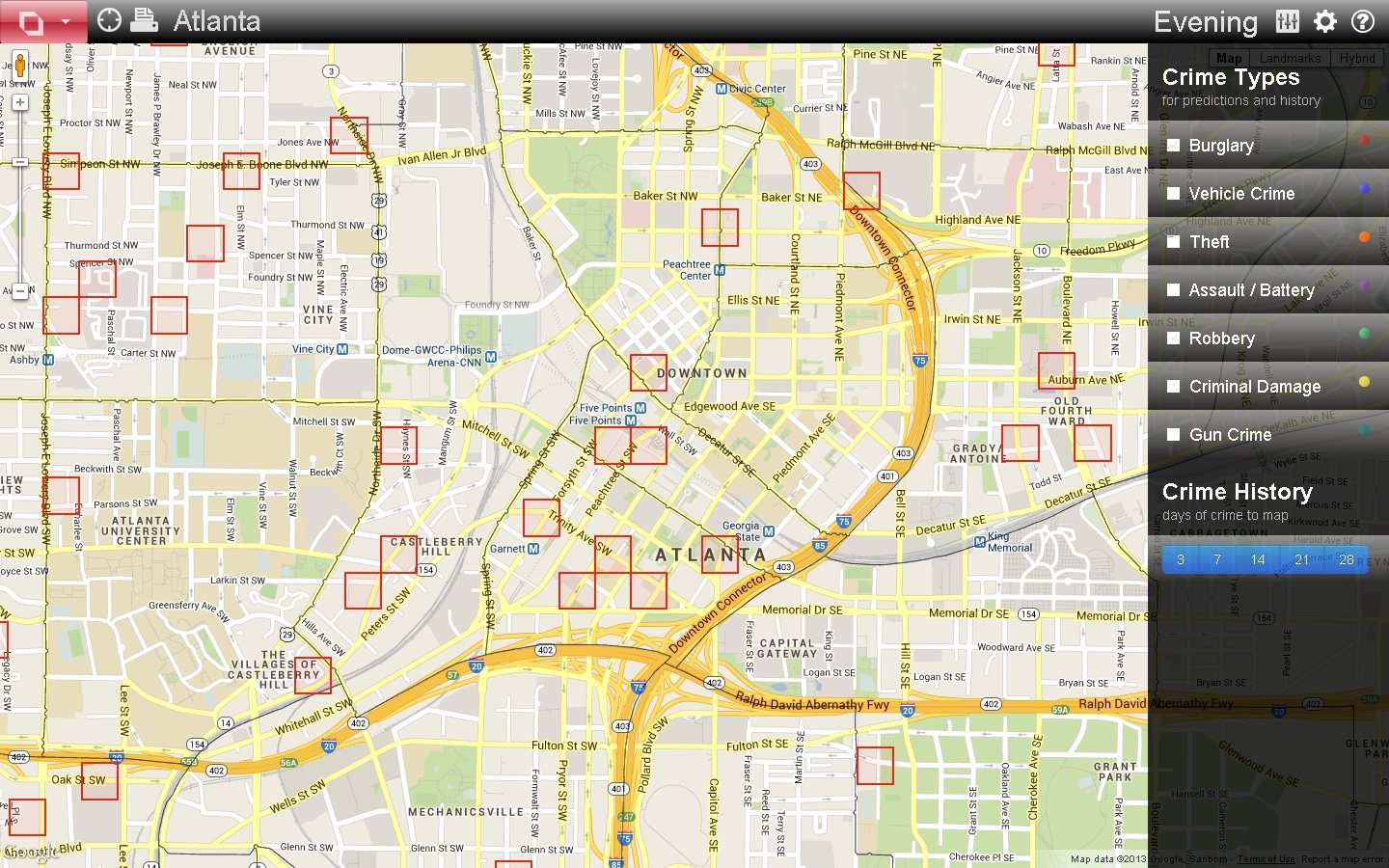}
  \caption{Screenshot from PredPol illustrating crime-prone hot spots. Pic from https://www.neoteo.com/predpol-una-plataforma-para-predecir-el-crimen/ }
  \label{fig:predpol}
\end{figure}

We briefly outline the historical context and evolution of hot spot policing and illustrate how it resonates with the task of identifying problematic polling booths. The shift of focus in policing from people (i.e., criminals) to places has roots in the deployment of heuristics such as {\it broken windows}~\cite{wilson1982broken} to identify crime-prone places. This was later mainstreamed -- most prominently under the mayorality of Rudy Giuliani in New York, whose regime was praised for effective place-based policing~\cite{langan20041} amidst scathing critiques focused on racial bias~\cite{noel2007blacks}. This was followed by strong scholarly arguments in favor of increased normalization of hot spot policing~\cite{sherman1995general}. Today, hot spot policing is embedded strongly within the umbrella of predictive policing pervasively within US and Europe through software such as Predpol (Geolitica\footnote{https://geolitica.com/}) and Palantir Gotham\footnote{https://www.palantir.com/platforms/gotham/}. An illustrative screenshot from PredPol has been included in Fig~\ref{fig:predpol}. We observe that predictive policing is built strongly on the principles of pre-crime~\cite{mcculloch2015pre} (the idea that the occurrence of a crime can be anticipated before it happens) and the primacy of {\it place} (as opposed to criminals, victims and temporality) in crime. These, we note, are precisely the premises upon which the idea of predicting problem polling booths are founded upon. Predictive policing techniques leverage historical patterns of crime and indicative cues, and this broad working philosophy makes their building blocks likely to be adapted by corporate software vendors to expand into the electoral AI market. 

Problematic polling booth prediction is a task that has its own nuances, and thus, bespoke methods for the same may leverage advances in related sub-disciplines of data-driven AI. For example, a pre-compiled set of tell-tale patterns that are indicative of impending polling place violence may be leveraged for automated discovery in time series streams, a task which may be cast within the framework of time series motif discovery~\cite{mueen2009exact}. 

\subsection{Extant or Potential Ramifications} 

Given the absence of extant usage of AI in problematic polling booth prediction, we focus on potential ramifications in using AI for the task in the future. Density gradients in policing of polling booths could have direct and indirect ramifications. As an example of a direct risk, observe that heavily policed polling booths could repel voters, reducing voter turnout; it was observed that police raids reduced turnout in Spain in 2017~\cite{rodon2018beaten}. Thus, it may be in the interests of the incumbent party to incentivize AI that designates areas where they expect lesser support as problematic, as a way to indirectly and silently sway electoral results in their favor. Besides this specific effect, there have been broader observations that policing of elections would influence electoral outcomes~\cite{verma2005policing}. 

Based on our conjecture that likely usage of AI for the task would reflect the patterns of usage of predictive and hot spot policing, the current understandings of risks within that realm provides a vantage point to ponder about the indirect risks for predicting problematic polling booths. It is also notable that the latter task, much like hot spot policing, is within the public sector, making learnings from hot spot policing likely pertinent for our task. Hot spot policing in particular, and {\it new policing} in general\footnote{a phrase used to refer to technology-driven policing}, has been argued to aggravate systemic racism within the American context~\cite{braga2019race} through subjecting minority neighborhoods to higher levels of policing. Other observations link an aggravation of police brutality to hot spot policing; for example, an article in the Hill\footnote{https://thehill.com/blogs/congress-blog/civil-rights/265795-police-brutality-is-not-invisible/} says: {\it `The epidemic of police brutality -- primarily affecting black males -- can be linked to the history of a technique called hot spot policing, ...’}. This has led to responses with skepticism even from the most vocal pioneers of the technology~\cite{weisburd2016does}. Applied AI has often been argued to have a performative (cf. descriptive) aspect, where they {\it reshape the very phenomenon they are supposedly modelling}~\cite{mcquillan2022resisting}. The observation of runaway feedback loops in predictive policing~\cite{ensign2018runaway} foregrounds this performative aspect of policing AI which, we surmise, could spill over in myriad ways to the use of similar technology in predicting problematic polling places. 

\subsection{Pathways Forward} 

We find it hard to envisage ways in which AI could be used in a reasonable manner to predict problematic polling places. Firstly, we observe that the nuanced nature of booth capturing makes it hard to predict using mathematical abstractions and statistical models. Secondly, given that AI cannot be expected to be foolproof, any errors made by AI in choosing problem polling booths would arguably have a needless indirect influence on electoral outcome. 

Yet, in cases where AI is used to prioritize police force deployment for polling booth protection, there could be ways to use it in transparent and principled ways to mitigate loss of confidence and trust among the public in the prioritization process. Since transparency and explainability have been discussed previously, we do not delve into those details again. We note that the task of spatial hot spot identification has been explored, in parallel to data mining methods, within the realm of statistical theory; a survey of related techniques appears in~\cite{deepak2016anomaly}. In particular, the {\it spatial scan statistic}~\cite{kulldorff1997spatial} has spawned a statistically principled family of methods. These, while being principled in deploying the usage of rigorous and bespoke statistical significance tests, suffer from poor scalability and efficiency, explaining their limited uptake within hot spot policing techniques. The statistical rigor in those methods would limit the extent of errors in predicting problematic booths. Further, there has been emerging literature on demographic fairness in hot spot detection~\cite{p2022fish}, with applications to hot spot policing. The overall idea is to ensure that the collective population across areas judged as problematic are demographically diverse and representative of the broader population. Attempts at blending transparency, statistical rigor and fairness principles in developing techniques for problematic polling place identification could mitigate some of the risks discussed earlier. 


\section{Voter Authentication}

\subsection{The Context} 

Ensuring that the vote is cast by legitimate voters is obviously of paramount importance to the integrity of the electoral process. Most electoral processes globally require some form of authentication at the polling place, such as a requirement that voters bring an identification document while visiting the polling place. While most focused studies have found very little evidence of impersonation~\cite{james2020electoral,ahlquist2014alien}, accusations of voter fraud are often used by political parties to discredit the electoral process and outcomes, to good media attention~\cite{fogarty2015news}. Such accusations exert pressure on democratic institutions to reinforce public trust in electoral processes through continuous attention to voter authentication. Within the context of online voting (e.g., Estonia), the usage of a digital id for voter authentication is protected by strong and robust encryption technologies\footnote{https://e-estonia.com/solutions/e-identity/id-card/}. Voter authentication within postal voting is multi-modal and implicit since the voter needs to make a request following which the voting package is sent to the registered address\footnote{https://en.wikipedia.org/wiki/Postal\_voting\_in\_the\_United\_States}. Our focus, in this section, is on the use of AI-based voter authentication methodologies at the physical location of the polling place. 

\begin{figure}
  \includegraphics[width=\columnwidth]{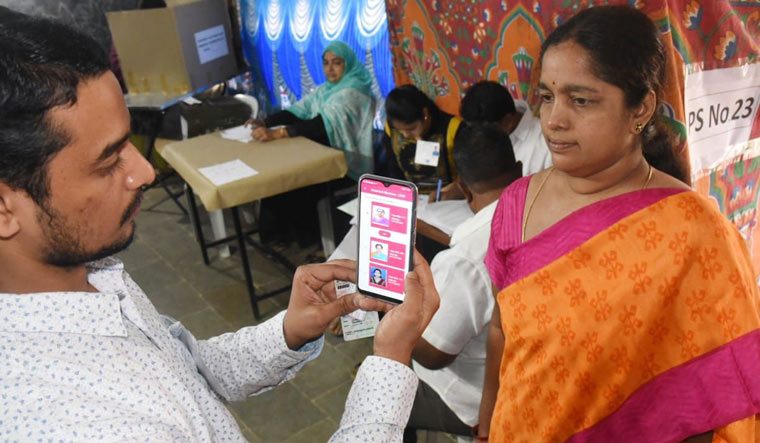}
  \caption{Usage of Face Recognition for 2020 Elections in Telangana (India). Pic from https://www.theweek.in/news/india/2020/01/23/India-first-poll-using-face-recognition-app-conducted-peacefully-in-Telangana.html }
  \label{fig:face-recognition-pilot-telengana}
\end{figure}

\subsection{Extant or Potential AI Usage} 

The use of AI for voter authentication, based on our literature search, is found to be very scarce. One of the notable usages has been within the context of a face recognition pilot in the 2020 Telangana municipal elections in India~\cite{allie2023facial}; a picture from a media report on the same appears in Fig~\ref{fig:face-recognition-pilot-telengana}. There have been academic prototypes focusing on video verification of voter identity~\cite{dunbar2015video}. However, this sparse uptake sits in very sharp contrast with AI literature which boasts of a plethora of person identity verification and fraud detection mechanisms using techniques such as face recognition~\cite{li2020review}, fingerprint recognition, iris recognition~\cite{mostofa2021deep}, retina scans, and counterfeit document detection~\cite{centeno2019recurrent,berenguel2019analysis}, among others. Given observations of improvements in document forging and an increasing realization that the promise of tamper-proof official documents remains a holy grail~\cite{baechler2020document}, it may be reasonable to expect a shift of focus from voter authentication based on identity documents to biometric based authentication. 

While it is clear that AI-based voter authentication is likely to see an enhanced interest in the times to come, a key aspect is whether there is an emerging data infrastructure that can enable the deployment of these technologies at the population scale. Most nations would have photos collected as part of voter enrolment, and thus have (potentially stale) national-level citizen photograph databases. These may not be sufficient, on their own, to implement and deploy facial recognition; it is in this context that deeper forms of national-level data collection gain importance. Considering the case of {\it India}, the largest democracy by population, there has been an emergence of biometrics-based database called {\it Aadhar} used as a gateway to access several services~\cite{rao2019aadhaar}. Aadhar records fingerprints, iris scans and photographs of each citizen, and boasts of a coverage of 95\% of the population\footnote{https://government.economictimes.indiatimes.com/news/digital-india/95-of-people-have-aadhaar-and-use-it-once-a-month-on-average-report/72236213}. Aadhar is routinely used to access public services through fingerprint scanners. While Aadhar is separate from the voter's identity card, there has been a recent initiative to link voter cards with Aadhar\footnote{https://en.wikipedia.org/wiki/Aadhaar\#Drive\_to\_link\_Aadhar\_with\_Voter\_ID\_card} which increases the readiness to use biometric authentication in elections. There has been much enthusiasm in academic circles to design Aadhar based biometric and digital (e.g., one-time-password) based authentication mechanisms in elections~\cite{roopak2020electronic,jain2023blockchain}. While Aadhar is significant due to the scale, other biometric-based population databases have emerged in recent times, of which Indonesia's e-ktp~\cite{darwis2011design} system is an example. These indicate that AI usage for voter authentication may be expected to be piloted widely in the near future. It also needs to be mentioned that there is significant resistance to national ID cards in the West, and protests have caused large-scale biometric based ID projects to be shelved to postponed (details in~\cite{ramakumar2010unique}). 

\subsection{Extant or Potential Ramifications} 

We now consider the myriad risks of using AI-based authentication mechanisms for voter identification. {\it First}, the issue of race and gender bias of facial analysis software was highlighted to much media attention in 2018~\cite{buolamwini2018gender}. It was shown that intersectional groups (e.g., black women) suffer very high rates of error (up to 34\%), whereas light-skinned males often record accuracies in excess of 99\%. To put this in perspective, observe that this could lead to significant numbers of women, minorities and intersectional groups wrongfully denied entry through the AI-based authentication mechanism (to fall back to manual authentication processes), inconveniencing and alienating them from the electoral process. {\it Second}, beyond facial analysis, biometric recognition systems such as fingerprint, iris and retina-based ones have seen very limited experimental analyses on possible bias in behavior. Fingerprint systems, the more popular among the above, have been observed to exhibit some amounts of demographic bias~\cite{marasco2019biases}. Fingerprint based authentication in Aadhar has been reported to have high prevelance of authentication errors among manual labourers in a small-scale survey~\cite{bhatti2012aadhaar}; large-scale studies need to be done to confirm whether these are chance occurrences or systematic bias against labourers whose fingerprints may have faded beyond the margin of error (as reported in the case of beedi workers\footnote{https://pulitzercenter.org/stories/fading-fingerprints-beedi-workers-india}). {\it Third}, there is a risk that usage of enhanced technology could reduce voter turnout, especially among marginalized communities. The face recognition pilot in Telangana (India) in 2020 recorded statistically significant reductions in voter turnout, amounting to more than $6$ percentage points. The study analyzing this~\cite{allie2023facial} outlines three potential explanatory reasons; logistical issues, shifts in fraudulent activity and increased alienation of marginalized religious communities (in this case, Muslims). The last concern is of significant importance and points to the potential embedded within such technologies to enhance the power of the state apparatus to engineer electoral calculus. {\it Fourth}, there are broader concerns with the technological solutionism (the belief that every problem can be solved by technocratic means) embedded within the urge to use AI-based biometric authentication, as outlined in~\cite{gelb2019biometric}. The article says that biometric elections need not necessarily enhance credibility and fairness of elections, and in some cases could even undermine it. Further, the enhanced resourcing diverted to maintaining up-to-date biometric authentication for election usage could be used to deepen surveillance towards undermining individual freedoms and political rights~\cite{hosein2013aiding}. 

\subsection{Pathways Forward} 

Much like the case of polling booth protection, we find it hard to envision a reasonable usage of AI for voter authentication, especially in the case of biometric-based authentication. Yet, we deliberate on the possible pathways to mitigate risks in the usage of AI-based biometric authentication of voters. {\it First}, we observe that non-technological administrative safeguards ought to be instituted to ensure that biometric authentication does not alienate (sections of) voters, through instituting other parallel options. As recommended in~\cite{bhatti2012aadhaar}, biometrics should not crowd out other authentication options and become a de facto compulsory alternative. {\it Second}, extensive auditing of biometric identification using carefully curated benchmarking (e.g., FRVT~\cite{grother2019face}) could help reveal biases to help direct policy and technological effort towards addressing them to mitigate biased operation. However, these have been warned as creating a false sense of fair AI, and concealing deeper concerns~\cite{raji2020saving}, something to be watchful of. {\it Third}, the unknown nature of the biases embedded within AI-based biometric authentication (since they have not been widely deployed at population-level) is a risk that deserves significant attention, towards developing inclusive authentication mechanisms. We note that many biases stem from the attention to the {\it typical} user in technology design, an ethos that stems from the commercial development of AI. AI development often makes use of {\it personas} (a popular product design tool~\cite{pruitt2010persona}), involving characterizations of typical users. The usage of personas ensures that the technology works well for the typical users, an important criteria for commercial products. However, within an election-focused authentication AI, the core interest is in ensuring that nobody is left out, which could be operationalized by focusing on identifying failure-prone `edge' cases, and making sure that the technology works for them too. Such edge-case thinking~\cite[Ch. 3]{wachter2017technically} could help foster higher levels of inclusivity and fairness in AI-based authentication. 

\section{Video Monitoring of Electoral Fraud}

\subsection{The Context} 

Monitoring elections for proper conduct is seen as an important measure to ensure free and fair elections. In particular, as noted in~\cite{hyde2014information}, monitoring ensures that there are ways to verify or vet post-election allegations of electoral fraud. The most common monitoring is that by independent international observers, but there could be bodies that monitor domestic elections, such as the Free and Fair Election Network (FAFEN\footnote{https://fafen.org/}) in Pakistan. Independent of the presence of electoral observers, electoral bodies may try to further electoral transparency through modern monitoring technologies such as CCTV/surveillance cameras, which have been seeing increasing global role in surveillance systems. CCTV systems have been argued to be more effective than monitoring by human observers while also serving to deter electoral fraud~\cite{obeta2021credible}. In contrast to electoral monitoring by observers, CCTV based monitoring, given its inherent data-oriented nature, enhances the role that AI can play in election monitoring, which is what makes this a topic of interest for this paper. The emergence of scalable AI that can identify events in real-time have been increasingly used in other contexts such as institutional security~\cite{kakadiya2019ai}, home alert systems~\cite{liang2021smart}, and health-motivated monitoring of elder citizens~\cite{huang2018video}; this makes the video-based election monitoring context a fertile avenue for AI usage.

\subsection{Extant or Potential AI Usage} 

\begin{figure}
  \includegraphics[width=\columnwidth]{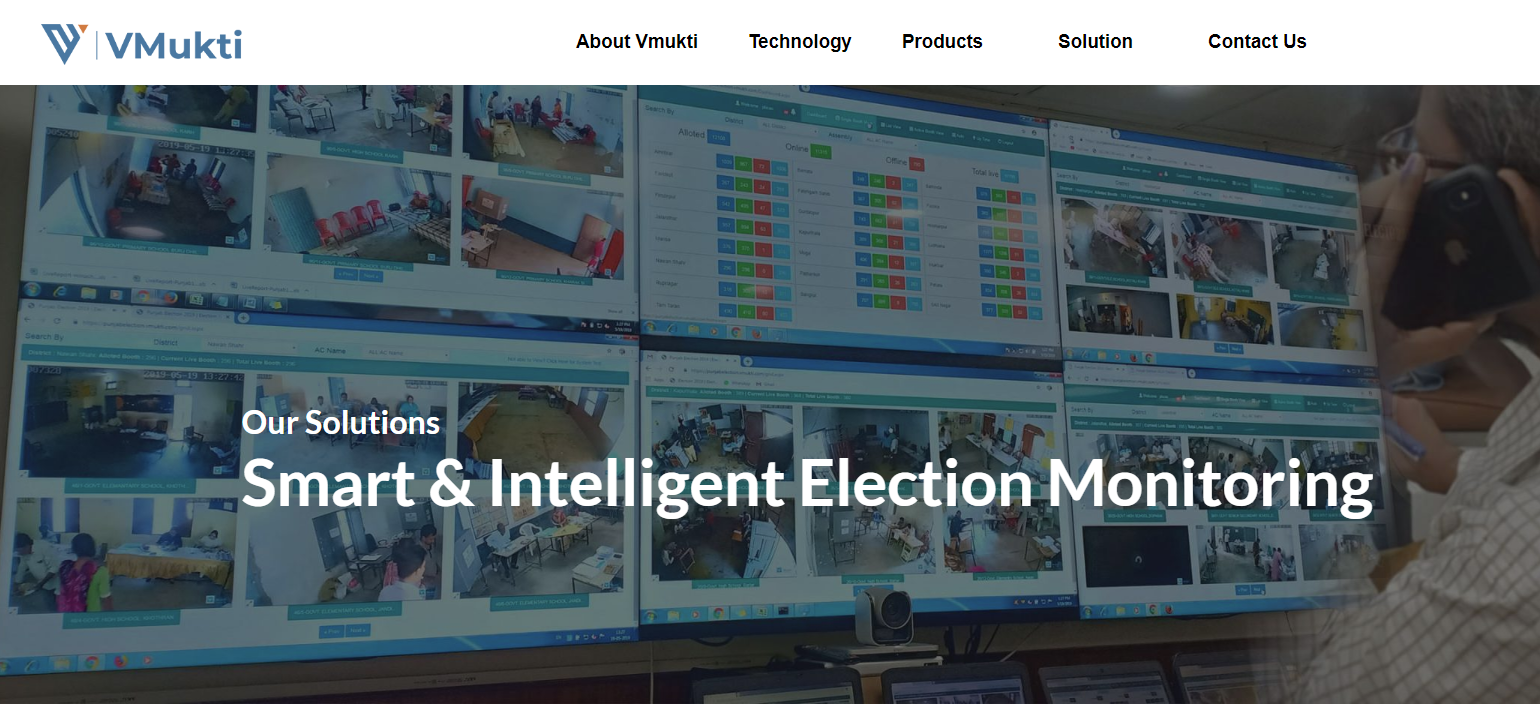}
  \caption{Page from VMukti, describing an election focused CCTV monitoring offering. Pic from https://www.vmukti.com/solution/election-surveillance/ }
  \label{fig:vmukti}
\end{figure}

While there has been significant deployment of CCTV based infrastructure in polling places globally, e.g., Russia~\cite{russiacctv} and India\footnote{https://theprint.in/politics/uttarakhand-cctv-monitoring-at-polling-booths-security-increased-for-polling-day/830178/}, there have been largely used to illustrate transparency and deter electoral fraud, or for citizen monitoring of elections. There is very limited public information available about how AI has been used over CCTV data streams, either in real-time or for post-election analyses. Among evidence of extant AI usage, an invited talk at an AI for elections workshop in 2021~\cite{gupta2021indian} mentions usage of AI and video analytics to verify number of votes counted by analyzing data from CCTVs within Bihar state elections in India. A news article~\cite{ahaskar2021officials} throws more light into the usage of AI within the Bihar polling process. It suggests that video captures of the Electronic Voting Machine is performed and analyzed using OCR technology to do a technology-based parallel counting of votes much before the actual post-election vote counting process. The article also suggests real-time usage of AI-based analytics to identify any discrepancy on vote counts to alert polling officials. The proprietary technology, codenamed JARVIS\footnote{https://www.staqu.com/\#what\_jarvis\_is}, is seen to be technology that is used for a variety of applications, and not a bespoke technology stack targeted at polling analytics. This is illustrative of how extant AI-based video analytics technology could transfer over to usage in polling booths, which forms our focus on the discussion of potential AI usage. It is also notable that there is an emerging prevalence of election-targeted video analytics solutions e.g., VMukti\footnote{https://www.vmukti.com/solution/election-surveillance/} as illustrated in Fig~\ref{fig:vmukti}. 

\begin{figure}
  \includegraphics[width=\columnwidth]{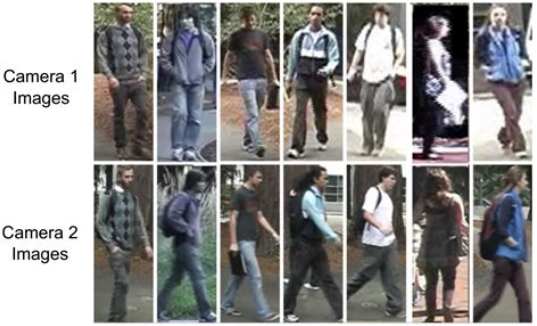}
  \caption{Illustration of Person Re-identification across two cameras. Pic from Rapid-rich Object Search Lab, NTU (SG) }
  \label{fig:reidentification}
\end{figure}

AI-based video analysis technology is most mature for the context of public safety, where it is largely developed for usage over data streams from surveillance cameras in public places. In 2016, it was reported that the average UK resident is captured across 70 cameras daily\footnote{https://www.cctv-surveillance.co.uk/news/uk-person-cctv-cameras-70-times-day-research/}, indicating the pervasive usage of video monitoring in contemporary societies. While we cannot summarize the extensive body of work on surveillance video analytics for public safety within this section, we provide some representative examples prioritized on their potential usage within the context of video-based electoral monitoring. One of the prime usages in public safety is towards identifying interesting events, along with their {\it what}, {\it when} and {\it where} of each event~\cite{zhang2019edge}. Certain types of events, such as violence~\cite{ramzan2019review}, may be directly useful for usage over video streams from polling booths. More general AI for anomaly detection over videos~\cite{sultani2018real} could also be leveraged as-is in polling booths to filter video streams for downstream manual analysis. It is conceivable that supervised ML models developed for public safety could deliver much higher accuracy for the polling booth setting, given the controlled and predictable environment of the polling booth. Of late, there has been a significant surge of interest in person re-identification~\cite{ye2021deep}, the task of tracking the same person across multiple video streams (e.g., tracking a criminal's movements around the city captured across a plurality of surveillance cameras), as illustrated in Fig~\ref{fig:reidentification}. Person re-identification is well-suited in developing a technological solution against multiple voting\footnote{https://www.aec.gov.au/About\_AEC/Publications/backgrounders/fraud-and-multiple-voting.htm}, a well-understood electoral fraud where the same person casts votes multiple times, violating the equal representation principle\footnote{https://en.wikipedia.org/wiki/One\_man,\_one\_vote}. 


\subsection{Extant or Potential Ramifications} 

The ramifications of usage of AI-based video monitoring of elections are numerous. {\it First}, the usage of technologies like JARVIS (referenced earlier) that integrate OCR with video processing to enable fine-grained electoral monitoring and even parallel candidate-specific vote counting pose serious questions relating to electoral integrity. Given that the voter and thus their identity is also captured, this could be seen as a pathway towards an explicit and apparent violation of secret suffrage\footnote{https://en.wikipedia.org/wiki/Secret\_ballot} ({\it aka} secret ballot or secret voting), a core principle of voting systems enshrined within the UN Declaration of Human Rights (Article 21.3\footnote{https://www.humanrights.com/course/lesson/articles-19-25/read-article-21.html}). The usage of proprietary software for monitoring and analytics risk leaking sensitive electoral information to private parties\footnote{Even if the analytics is hosted within state-owned infrastructure, some data-based feedback, at least in the form of errors detected, would need to be passed on to the private vendors to build the next generation of the solution.} outside the remit of the electoral bodies who bear sole responsibility for the conduct of the elections, which is another risk to electoral integrity. Without going into details and consequences, we note here that real-time candidate-specific vote counting could pose very grave consequences to electoral integrity by creating privileged information that could violate expectations of transparency in vote counting\footnote{https://followmyvote.com/importance-of-transparency-in-voting/}. {\it Second}, usage of video surveillance technology, as observed in the case of face recognition~\cite{allie2023facial}, has the potential to alienate marginalized communities. An emerging trend of `protest surveillance'\footnote{https://www.cnbc.com/2020/06/18/heres-how-police-use-powerful-surveillance-tech-to-track-protestors.html} employs these very same technologies (e.g., video surveillance and analyses); against this backdrop, marginalized communities that engage in protests may find electoral surveillance much more alienating. There is also a broader context of studies on how contemporary surveillance technologies are informed by the long history of racial formation and policing of black lives in the US~\cite{browne2015dark}, which highlights that there could be nuanced implications of video monitoring on marginalized communities. {\it Third}, as a minor point, we observe that mainstreaming of video-based surveillance may undermine other forms of more procedurally rigorous election monitoring methods such as in-person visits by independent observer panels. 

\subsection{Pathways Forward} 

As outlined earlier, one of the fundamental risks of fine-grained video monitoring of elections is the threat it poses to electoral integrity principles such as secret suffrage and transparency of vote counting. This is in addition to enhanced alienation of groups of voters brought about by the presence of the surveillance technology itself. The main purported benefits, on the other hand, are to deter electoral violence and fraud while effectively detecting unusual activity that merits manual attention. Given this context, the pathways in deploying technological election monitoring, if such solutions need to be pursued, should ideally negotiate a space that is compatible with electoral integrity while realizing the benefits of detecting electoral violence and other unusual activities. {\it First}, a pathway towards complying with principles of electoral integrity would be to institute only shallow forms of monitoring in lieu of full-blown video monitoring. Pathways within this stream could involve using audio monitoring, fine-grained motion detection sensors, or very low-resolution cameras (that may capture only blurred images which can help identify movements, but not individual people or their identities), or a surveillance setup involving some or all of the above. Such shallow monitoring would not allow packaging the solution with technologies such as OCR (recollect JARVIS) or face recognition since those would not work with blurred video streams, mitigating the threat to electoral integrity. On the other hand, bespoke computer vision and pattern recognition technology would need to be developed in order to ensure effective violence and anomalous events detection over such streams, leading to new directions in AI research. If multiple sensing modalities (e.g., audio, motion, blurred videos) are involved, technologies for correlating evidences across such multi-modal streams would need to be developed to ensure that the benefits of the system are realized to a reasonable extent. As may be obvious, a reasonable first step towards enabling or kickstarting this direction would be legislative or regulatory viz., a legal regulation disallows the use of full-blown video monitoring in jurisdictions where they are currently not outlawed. {\it Second}, shallow monitoring comes with a challenge, that of ensuring that citizens understand and appreciate that such monitoring would not allow identification of individuals. This trust issue, while a non-technical challenge, needs to be resolved satisfactorily in order to mitigate the risk of alienating citizen groups and marginalized communities. One solution path would be through {\it open data} initiatives, whereby all such monitoring data is made publicly available through (potentially real-time) web streams, to illustrate the shallowness of the monitoring in the open. This could encourage hacktivism based experiments, which could highlight any issues, opening up a pathway towards continuous auditing and feedback-based improvement of the shallow monitoring solutions. Open data could also encourage citizen audits of elections; a notable case comes from Russia where citizen analyses of CCTV footages foregrounded instances of ballot stuffing\footnote{https://www.huffpost.com/entry/russia-election-2012-ballot-box-stuffing\_n\_1321379}. 

\section{Visibility of AI Usage}

We now turn our attention to the {\it visibility} of AI usage to the voters and the general public, an important consideration. As an example, usage of technological voter authentication (e.g., using face recognition) is explicitly visible, whereas the usage of AI in voter list cleansing is likely visible only when process documents are available in the public domain. Visibility is an important consideration since visibility ensures public scrutiny, making low visibility usage more risky. In other words, usage of AI in a medium risk manner with very low visibility could potentially have worse long-term consequences than high risk high visibility AI usage. We briefly review our five avenues of interest using the perspective of analyzing visibility of (extant or potential) AI usage. 

Voter List maintenance is naturally a back-office task which is carried out by election officials in between elections, with a usual surge in activity in the run-up to elections to ensure election readiness. Given that this is a back-office task, usage of AI within it is not directly visible to the public. Regulatory constraints that may require that a reason be given to voters from whom confirmations are sought may still not expose the AI usage, given the vast literature of post-hoc explainable AI. Post-hoc explainable AI methods such as LIME~\cite{ribeiro2016should} are capable of providing an automated explanation for any decision (in this context, the decision by a classifier to choose to seek confirmation from a particular voter), making it easy to confirm to regulatory constraints on explanations without exposing the usage of AI. Unless there is a regulation that the use of AI within decision making needs to be transparent (or unless these fall under right to information regulations), the visibility of usage of AI within voter list maintenance could be very limited. 
                                        
The usage of AI to determine polling booth locations and identifying problem booths, as in the case of voter list maintenance, could also be done under very limited visibility. Given that polling booth determination and problem booth identification involve much fewer decision making instances (of the order of the number of polling booths) and further manual inspections are necessary to arrange polling booth logistics or deploy additional patrols, this would likely have a substantive human component in the decision making pipeline. This makes any AI inputs into the process quite opaque. 

In contrast to the above discussed avenues, voter authentication is an arena where AI usage would inevitably have high visibility. The usage of biometrics for authentication would require polling officials to procure equipment and inform voters, so that they can be effectively deployed. In the case of facial recognition, most current algorithms would require capturing stable photos, as seen in Fig~\ref{fig:face-recognition-pilot-telengana}. If the technology matures to an extent where CCTV footages can be used for face-recognition based voter authentication, the visibility of AI usage may reduce. However, as of now, the usage of AI within the task of voter authentication is accompanied by high visibility to voters. 

Video monitoring, the fifth among our chosen avenues, involves the usages of cameras and connectivity infrastructure. This makes it a high visibility usage of AI since the observant voter would likely easily spot its usage. The emergence of small and invisible spy cameras pose a risk; however, we believe it is unlikely that electoral bodies would be keen on using such `tricks' to reduce visibility. 

\begin{table*}
\begin{center}
\begin{tabular}{|c|>{\centering}p{4cm}|>{\centering}p{4cm}|>{\centering}p{4cm}|}
    \hline
    {\bf Avenue} & {\bf Technology Readiness} & {\bf Risk Level} & {\bf Visibility of AI Usage} \tabularnewline
    \hline
    \hline
    Voter List Maintenance & \cellcolor{blue!60}HIGH & \cellcolor{red!30}MEDIUM & \cellcolor{green!20}LOW \tabularnewline
    \hline
    Polling Booth Locations & \cellcolor{blue!30}MEDIUM & \cellcolor{red!30}MEDIUM & \cellcolor{green!10}VERY LOW \tabularnewline
    \hline
    Predicting Problem Booths & \cellcolor{blue!60}HIGH & \cellcolor{red!60}HIGH & \cellcolor{green!10}VERY LOW \tabularnewline
    \hline
    Voter Authentication & \cellcolor{blue!100}VERY HIGH & \cellcolor{red!60}HIGH & \cellcolor{green!100}VERY HIGH \tabularnewline
    \hline
    Video Monitoring & \cellcolor{blue!100}VERY HIGH & \cellcolor{red!100}VERY HIGH & \cellcolor{green!50}HIGH \tabularnewline
    \hline
\end{tabular}
\caption{Overall Summary of Key Assessments. \label{tab:summary}}
\end{center}
\end{table*}

\section{Overall Summary}

We now summarize the discussion so far to provide a quick high-level view of our assessments and impressions. Table~\ref{tab:summary} provides an overview of our evaluation of each of the five considered avenues in terms of technology readiness, risk level and visibility of AI usage. While these are not objective assessments, but based on an information evaluation of the current state of play, these are to be taken with caution. Yet, we illustrate the rationale behind some assessments in the interest of exposing our thought process. AI technology that is aligned with determining polling booth locations is available across different streams, but we assess that such technology would need non-trivial adaptation for usage in the task; this reasoning leads us to evaluate the technology readiness to {\it medium}. On the other hand, hot spots policing is more or less readily usable for the task of identifying problem booths leading to a technology readiness assessment of {\it high}. The risk level of using AI in voter list maintenance and polling booth determination is assessed to be {\it medium} due to the virtual inevitability of using them only in a human-in-the-loop manner, ensuring some amount of expert oversight. In contrast, the usage of video monitoring, as we saw earlier, could sit in sharp tension with principles of electoral integrity; this is acknowledged through the {\it very high} risk level in the table. Given the discussion of visibility in the preceding section, we do not elaborate on that here to avoid repetition. 

\section{Other Avenues of AI Usage}

While our choice of five avenues of AI usage was intended to give a general picture of the horizons in this area, there are obviously other aspects within core electoral processes where AI could play a role. We briefly consider a few such areas, to serve as some directions for future reading for the interested reader. 

{\it First}, there are certain areas where peripheral electoral processes (recollect our core-peripheral distinction) could feed into the core processes. One such arena is that of opinion polls and exit polls. While opinion polls date back to two centuries~\cite{tankard1972public}, exit polls have been used for half-a-century~\cite{moon2022exit}. These have become regular permanent and important fixtures that the public care about, especially for elections in Western liberal democracies. Of late, with the failures in predicting the Brexit referendum and the 2016 US elections\footnote{https://www.scientificamerican.com/article/why-polls-were-mostly-wrong/}, popular expectations on accuracies of pollsters' numbers may not be at a high. However, if confidence in such polls rebound, there is potential for narratives that view them as ground truth, something that the actual result ought to align to. This could deliver fertile ground to use deviations between actual results and opinion polls to determine which booths or areas should be subject to recounting. There are extant technologies that are aligned to this task~\cite{Fish2017On}. In certain electoral systems such as Mexico~\cite{mendoza2016quick} and France\footnote{https://www.thelocal.fr/20220407/how-do-the-french-produce-such-accurate-early-election-results/}, the election authorities use statistical sampling to provide a quick preliminary result. Similar quick-counts or parallel vote tabulation are growing in popularity in Africa\footnote{https://en.wikipedia.org/wiki/Quick\_count}. Technocratic solutions for selective recounting could propose usage of deviations between the actual count and such quick counts (in lieu of or in addition to opinion/exit polls), to determine which areas or booths should be targeted for recounting. As argued in~\cite{aparicio2006fraud}, such divergence-based determination of selective recounting could be biased, and random recounts are arguably safer. 

{\it Second}, the COVID-19 era has seen an increased usage of mail-in ballots, especially so during the 2020 US Presidential elections when 65 million voters exercised their franchise through mail-in ballots\footnote{https://www.pewresearch.org/fact-tank/2020/11/10/most-mail-and-provisional-ballots-got-counted-in-past-u-s-elections-but-many-did-not/}. The National Vote at Home Institute, a postal ballot advocacy group, found major deficits in infrastructural readiness for large-scale mail voting~\cite{voteathomepolicy}. These include 15 states that lack steps to verify voter addresses, and 17 states that do not mandate a signature verification process. Within many states, it was reported that signature mismatches could lead to disqualifying the vote\footnote{https://www.latimes.com/politics/story/2020-10-13/signature-verification-rules-lawsuits} without the provision of a chance to fix the mistake. It has been pointed out~\cite{kayserbril2022algorithmic} that many US states use machine learning for signature verification. It has been also noted therein that machine learning based signature verification systems may not have a high enough accuracy rate, and that their usage could be a source of disenfranchisement with a demographic gradient since most systems are trained on native English speakers. It is, however, not clear as to whether mail-in voting would still continue to be used on a large scale beyond COVID-19; however, if they become institutionalized, AI-based signature verification is an avenue of potentially intense activity.

{\it Third}, social media has come to be recognized as a major player in elections, especially with the surfacing of the Cambridge Analytica scandal~\cite{hinds2020wouldn}. The scandal itself relates to supercharging electoral campaigns, one which we regard as outside the core electoral process. However, the role of Cambridge Analytica (CA) in Nigeria (2015)\footnote{https://www.theguardian.com/uk-news/2018/mar/21/cambridge-analyticas-ruthless-bid-to-sway-the-vote-in-nigeria} and Kenya (2017)\footnote{https://advox.globalvoices.org/2017/11/03/data-and-democracy-what-role-did-cambridge-analytica-play-in-kenyas-elections/} ignite much serious concerns. In both cases, CA worked in favor of the incumbent leaders, which raise concerns about whether they had access to privileged government information, especially relating the personal financial and medical records of opposition leaders, as noted in the Guardian article above. Broadly, CA's systems were later found to have actively interfered in electoral processes and worked against social cohesion~\cite{nyabola2018digital,mohamed2020decolonial}. Broadly, access to privileged information could provide unforeseeable pathways for such tech giants to influence core electoral processes. In addition, the use of social media insights to inform decision-making within core electoral processes provides another facet of potential AI-driven interference. 

\section{Discussion}

Our focus so far has been to show how what we regard as core electoral processes could be seen as fertile ground for AI usage, to illustrate the risks that such usage may pose, and to initiate deliberations around pathways that could mitigate the risks. We now consider some broader contexts. 

The period after the 2008 financial crisis saw the emergence of austerity, or slashing public spending, as an attractive proposition~\cite{mcquillan2022resisting}. This has arguably fueled the uptake of AI within governments, as a new form of automation that will enable making newer forms of cuts to public spending. It is in this context that one would need to examine whether AI is simply automation, or something beyond. Gigerenzer~\cite[Ch. 4]{gigerenzer2022stay} illustrates how AI-based decision making has a character that is sharply different from human decision making. This has been long known, though hardly spoken of in that way, through the nature of adversarial attacks such as one-pixel attacks~\cite{su2019one}; one-pixel attacks vary an image by just one pixel, a change that a human eye can barely make out, to affect a change in the decision made by the classifier. This difference in the nature of decision making reflects in the fact that humans and AI make very different forms of errors in undertaking the same task. This resonates with the observations in~\cite{mcquillan2022resisting} that usage of AI alters the form of decision making. Given the enthusiasm towards AI usage within governments, such AI adoption is done without any democratic debate. This amounts to altering the social contract structurally without transparency or democratic deliberations. Radical structural changes to social contracts within algorithmic governance have been noted in UK welfare policy under the veneer of `total transformation' and the `digital welfare state' in the post-Brexit era~\cite{alston2018statement}. AI as a means of altering laid-out processes latently without democratic debate could cause significant changes within the realm of electoral processes, with far-reaching consequences to the foundations of democracy. 

There have been observations on the increasing corporatization of AI research~\cite{ahmed2020democratization}, with trends pointing to an increasing share of industrial research featuring within the top AI avenues in recent years. While significant AI research has been taking place within public universities, governments have been increasingly relying on private solution providers for AI usage within core functions, such as policing and migration~\cite{naranjo2019privatization}. Even if the process responsibility is still held within the public sector, the AI building blocks are often very sophisticated and opaque that public sector employees may not be able to meaningfully audit suggestions from the AI. It has been observed that transparency and a critical audience is key to algorithmic accountability~\cite{kemper2019transparent}. Given this scenario, the usage of AI thus implicitly leads to a form of privatization of public sector decision-making, where not just the ownership, but the agency of decision-making is outsourced to privately developed and owned AI. As observed in~\cite{mittelstadt2016ethics}, algorithms are inescapably value-laden; this leads to the current trends in AI uptake providing a channel for corporate values to leech into public sector decision making. Decision making within electoral processes making movements within the spectrum of values should be regarded as being of significant concern. 

Overall, we observe that there are challenges, risks and concerns at various levels when it comes to AI usage within core electoral processes. This points to the need of more debates and studies into the topic, and the importance of ensuring that regulatory boundaries are set in very careful but unambiguous ways. 

\section{Conclusions}

We considered the context of increasing enthusiasm on AI usage in contemporary societies, and the differentiated uptake across the public and private sectors. Drawing upon observations of increasing AI usage within the public sector, we outlined the boundaries of our analysis as that focusing on usage of AI within core electoral processes viz., those that relate to the time, place and manner of elections. This, we observed, sits in contrast with the general focus on AI usage within the electoral information ecosystem (e.g., campaigns, social media). We then enumerated five avenues within core electoral processes which currently witness or are amenable to AI uptake viz., voter list maintenance, determining polling booth locations, predicting problem booths, voter authentication and video monitoring of elections. Through a detailed analysis of each of these avenues, we outlined the potential pathways of AI usage and the ramifications they could bring about. For each avenue, we also identified potential directions to mitigate the risks, some of which point to new directions of AI research. We considered visibility of AI usage as an important consideration in mapping the risks that AI usage poses to core electoral processes. We then went on to profile the five avenues on the basis of technology readiness, risk levels and visibility. This was followed by a listing of other potential avenues of AI usage, which are potentially useful to the interested reader who is seeking to explore this area. We then discussed some overarching issues concerning broader issues of AI uptake in the public sector. As may be obvious to the reader, the usage of AI within core electoral processes is a volatile and interesting space which is likely to see significant activity in recent years. We hope our attempt at mapping the horizons in this arena will help deepen the debates on the topic. 

\section*{Acknowledgement}

Stanley Simoes has received funding from the European Union’s Horizon 2020 research and innovation programme under the Marie Skłodowska-Curie grant agreement No 945231; and the Department for the Economy in Northern Ireland.

\bibliographystyle{aaai}
\bibliography{refs-corrected}

\end{document}